\documentclass[onecolumn,prc,showpacs,preprintnumbers,12pt] {revtex4-2}
\usepackage{graphicx,latexsym,amssymb,amsmath,color,multirow,mathrsfs,booktabs,ifpdf,epsfig,dcolumn,bm,longtable}
\usepackage{changes}
\usepackage{hyperref}
\hypersetup{colorlinks=true,linkcolor=blue,filecolor=magenta,urlcolor=cyan}
\date{\today}

\begin{document}
\title{Low-energy $^7$Li($n,\gamma$)$^8$Li and $^7$Be($p,\gamma$)$^8$B radiative capture reactions within the Skyrme Hartree-Fock approach}
	\author{Nguyen Le Anh}
	\email{anhnl@hcmue.edu.vn}
	\affiliation{Department of Physics, Ho Chi Minh City University of Education, 280 An Duong Vuong, District 5, Ho Chi Minh City, Vietnam.}
	\affiliation{Department of Theoretical Physics, Faculty of Physics and Engineering Physics, University of Science, Ho Chi Minh City, Vietnam.}
	\affiliation{Vietnam National University, Ho Chi Minh City, Vietnam.}

	\author{Bui Minh Loc}
	\email{buiminhloc@ibs.re.kr}
	\affiliation{Center for Exotic Nuclear Studies, Institute for Basic Science (IBS), Daejeon 34126, Korea.}

\begin{abstract}
The electromagnetic dipole transitions in $^7$Be($p,\gamma$)$^8$B and $^7$Li($n,\gamma$)$^8$Li reactions at the keV-energy region were analyzed simultaneously within the Skyrme Hartree-Fock potential model. The Skyrme Hartree-Fock calculation is adopted as a microscopic approach to obtain consistently the single-particle bound and scattering states in the calculation of the radial overlap function within the potential model. All non-resonant and resonant electromagnetic dipole transitions are taken into account. The electric dipole transitions are successfully described with the slightest adjustment. The resonant magnetic dipole transitions at $633$ keV and $2184$ keV of $^7$Be($p,\gamma$)$^8$B reaction, and the one at $222$ keV of $^7$Li($n,\gamma$)$^8$Li are also analyzed. The astrophysical $\mathcal{S}_{17}(0)$ factor of $^7$Be($p,\gamma$)$^8$B reaction is found to be $22.3$ eV\,b.
\end{abstract}

\maketitle

\section{INTRODUCTION}
The nucleon radiative-capture reaction in which the atomic nucleus captures the incoming nucleon followed by the emission of electromagnetic radiation plays an important role in nuclear astrophysics. Both $^7$Li($n,\gamma$)$^8$Li and $^7$Be($p,\gamma$)$^8$B reactions are key processes in nuclear astrophysics. The $^7$Li($n,\gamma$)$^8$Li reaction is important due to its relation to the nucleosynthesis of heavy nuclei in the inhomogeneous Big Bang model \cite{kaj02}. While the $^7$Be($p,\gamma$)$^8$B is crucial in the production of high-energy solar neutrinos and the nucleosynthesis of low-mass stars \cite{ade98,ade11}. It is, however, one of the reactions with the largest uncertainty \cite{ade98,ade11} as it is required to be measured at extremely low energies to determine the cross section at the zero energy known as the $S_{17}(0)$ factor. 

The $^7$Li($n,\gamma$)$^8$Li and $^7$Be($p,\gamma$)$^8$B reactions were studied simultaneously in many previous works \cite{bar80,des94,bro96,ben99,nag05,bak06} due to the isospin symmetry in the $n+^{7}$Li and $p+^{7}$Be systems. For instance, the neutron optical potential for the $^{7}$Li target was used to reproduce the cross section for the reaction $^7$Be($p,\gamma$)$^8$B down to the very low energies \cite{bar80,bro96,ben99,nag05}. Therefore, $^7$Li($n,\gamma$)$^8$Li and $^7$Be($p,\gamma$)$^8$B are efficiently analyzed side by side.

Among the theoretical studies, there are a variety of approaches for the radiative-capture reaction such as the phenomenological potential model \cite{rob73, hua10, xu13, tur21}, folding model \cite{kit02, anh21NPA}, shell-model calculation \cite{ben99,ben00}, multi-cluster calculation \cite{des94,sco95,des04NPA,des04PRC} or phenomenological $R$-matrix \cite{bak95,bak06,pan19}. For the $^7$Be($p,\gamma$)$^8$B reaction, in particular, there were also the \textit{ab initio} calculations \cite{nav06,nav11} for the electric dipole ($E1$) transition and recently for the magnetic dipole ($M1$) transition included \cite{kra22}. A recent review was given in Ref.~\cite{des20}. The potential model is the simplest but powerful tool for the study of radiative-capture reactions as the capture process can be simply considered as the electromagnetic transition between the continuum state and the bound state. The radiative-capture reaction can be also approached with the potential model using the Skyrme Hartree-Fock calculation to obtain simultaneously the bound and scattering states \cite{anh21PRC103, anh21PRC104}. 

In the present work, the $^7$Li($n,\gamma$)$^8$Li and $^7$Be($p,\gamma$)$^8$B reactions were analyzed within the potential model using the Skyrme Hartree-Fock calculation. All electromagnetic dipole transitions are in the consideration of their contributions. In the case of the $^7$Li nucleus, the experimental data at low energies \cite{imh59,wie89,nag05,izs13} were reproduced efficiently both non-resonant $E1$ transition and resonant $M1$ transition at $222$ keV. In the case of the $^7$Be nucleus, all measured data \cite{kav69,vau70,fil83,kik98,ham98,str01,bab03,sch06,jun10} were well reproduced including not only both resonances at $633$ keV and $2184$ keV but also the non-resonant contribution up to $3000$ keV.

The results show that the single-particle scattering and the bound wave functions being the main inputs of the potential model are well-described using the two parameters of the approach. For the non-resonant $E1$ transition, only one parameter is adjusted. The spectroscopic factors that play a key role in the method are determined by the experimental data at a few hundred keV. It helps in determining precisely the $\mathcal{S}_{17}(0)$ factor that is 22.3 eV b in the present work.

The formulas for the $E1$ and $M1$ transitions are presented in the next section. The Skyrme Hartree-Fock calculation was described in detail in Refs.~\cite{anh21PRC103, anh21PRC104}. The section on results and discussions is started with the $E1$ transitions having no resonance followed by the $M1$ transition with low-lying resonances.

\section{METHOD OF CALCULATION}
\subsection{Electromagnetic dipole transition}
For the proton radiative-capture reaction, the energy-dependent astrophysical $\mathcal{S}$ factor $\mathcal{S}(E)$ is defined by
\begin{equation}
    \mathcal{S}_{aA}(E) = E\exp[2\pi\eta(E)]\sigma(E),
\end{equation}
where the Sommerfeld parameter describing the $s$-wave barrier penetration is $\eta(E) = Ze^2/(\hbar v)$ with $v$ being the initial relative velocity between the incident particle with atomic mass number $a$ and the target with the charge number $Z$ and the mass number $A$. In our case, $\mathcal{S}_{aA}(E)$ is the $\mathcal{S}_{17}(E)$. The general radiative-capture cross section $\sigma(E)$ is given in Refs.~\cite{bay83,hua10,xu13}. The electric quadrupole ($E2$) transition is negligible in the energy of interest within the potential model \cite{kim87,typ97,ber98,sch06,tur21}. Our present calculations are restricted to the $E1$ and the $M1$ transitions.

The initial state $| [I \otimes (\ell \otimes s)j]JM\rangle$ and the final state $| [I \otimes (\ell'\otimes s)j']J'M' \rangle$ of the system are assumed as the core that is the target nucleus with one additional nucleon that is the captured nucleon at a single-particle state. The core has the internal spin $\bm{I}$ that is unchanged in the calculation. Note that the ground states of the target nuclei are $I^\pi = 3/2^-$ in the study. The total relative angular momentum of the nucleon-target system is $\bm{j} = \bm{\ell} + \bm{s}$ with $\bm{\ell}$ being the relative orbital angular momentum and $s = 1/2$ for the nucleon. The channel spin of the initial system is $\bm{J} = \bm{I} + \bm{j}$. Therefore, the radiative-capture cross section in our case is expressed as \cite{rol73,hua10,xu13}
\begin{equation}\label{sigmaE}
    \sigma(E) = \dfrac{4}{3}\dfrac{1}{\hbar v} \left(\dfrac{4\pi}{3} k^3_\gamma \right) \dfrac{1}{(2s+1)(2I+1)} \sum_{\ell j J} \left( |\mathcal{M}_{E 1}|^2 + |\mathcal{M}_{M1}|^2 \right),
\end{equation}
where $k_\gamma$ is the wave number of the photon defined as $\hbar c k_\gamma = E + Q$ with $Q$ being the $Q$ value of the reaction. $\mathcal{M}_{\Omega 1}$ with $\Omega$ being $E$ or $M$ is the reduced matrix element of the electromagnetic dipole transitions. We herein follow the coupling scheme for angular momenta in Refs. \cite{kim87,ber03,hua10}.
\begin{equation}\label{Mstart}
    \mathcal{M}_{\Omega 1} 
    = \langle [I \otimes (\ell' \otimes s)j']J' || \mathcal{O}_{\Omega 1} || [I \otimes (\ell \otimes s)j]J\rangle.
\end{equation}

In the case of the $E1$ transition, the operator is given by 
\begin{equation}\label{OE1}
    \mathcal{O}_{E1} = C_e r Y_1^\mu,
\end{equation}
where $C_{e} = m(\tau-Z/A)e$ is the effective charge in which $e$ is the electric charge, $m$ the reduced mass, and $\tau=0$ for neutron and $\tau=1$ for proton; and $Y_1^\mu$ ($\mu = -1, 0, 1$) is the spherical harmonic function. The $\mathcal{M}_{E1}$ can be reduced to the calculation of the single-particle (s.p.) reduced matrix element \cite{sha13}
\begin{equation}\label{ME12}
    \mathcal{M}_{E1} = C_{e} S_F^{1/2}
    (-1)^{I+j'+J+1}
    \hat{J}\hat{J'}
    \left\{ \begin{matrix} 
        j' & J' & I \\ J & j & 1 
    \end{matrix} \right\}
    \mathcal{M}_{E1}^{(\rm s.p.)},
\end{equation}
where $\hat{l} = \sqrt{2l + 1}$ and the curly bracket is Wigner $6j$ coefficient. $S_F$ is the spectroscopic factor that is introduced to take into account the missing configurations of the system; in the other words, it is the fractional parentage coefficient that the system can be described as our assumption \cite{rol73,nag05}. It also contains missing correlations in the calculation of the single-particle state. Within the potential model for the radiative-capture reaction, its value is finally adjusted to reproduce the experimental data.

For the $M1$ transition, the operator is \cite{ber03,hua10,xu13,tur21}
\begin{equation}\label{OM1}
    \mathcal{O}_{M1} = \sqrt{\frac{3}{4\pi}} \left[C_m {\bm \ell_\mu} + 2\left(\mu_\tau {\bm s_{\mu}} + \mu_A {\bm I}\right)\right].
\end{equation}
With $\mu_N \approx 0.105e$ fm being the nuclear magneton, the effective magnetic moment is $C_{m} = m(\tau+Z/A^2) \mu_N$. The magnetic moments used in this work are $\mu_p = +2.793\mu_N$, $\mu_n = -1.913\mu_N$, $\mu_A = -1.399\mu_N$ for $^{7}$Be \cite{oka08} and $+3.256 \mu_N$ for $^{7}$Li \cite{bec74}. The reduced matrix element for the $M1$ transition is, therefore, written as the summation of three terms \cite{ber03,hua10,xu13,tur21}
\begin{equation}
    \mathcal{M}_{M1} = S_F^{1/2} \sqrt{\dfrac{3}{4\pi}} 
    \left( \mathcal{M}_{M1}^{(0)} + \mathcal{M}_{M1}^{(1)} + \mathcal{M}_{M1}^{(2)} \right). \label{M1}
\end{equation}
Similarly to the case of $E1$ transition in Eq.~\eqref{ME12}, we have
\begin{align}
    \mathcal{M}_{M1}^{(0)} &= C_m (-1)^{I+j+J'+1} \hat{J}\hat{J'} 
    \left\{\begin{matrix} 
        j' & J' & I \\ J & j & 1 
    \end{matrix} \right\}
    \mathcal{M}_0^{(\rm s.p.)}, \label{MM10}
    \\
    \mathcal{M}_{M1}^{(1)} &= \mu_\tau (-1)^{I+j+J'+1} \hat{J}\hat{J'} 
    \left\{\begin{matrix} 
        j' & J' & I \\ J & j & 1 
    \end{matrix} \right\}
    \mathcal{M}_1^{(\rm s.p.)}, \label{MM11}
    \\
    \mathcal{M}_{M1}^{(2)} &= \mu_A \delta_{jj'} (-1)^{I+j'+J+1} \hat{J}\hat{J'}  
    \left\{\begin{matrix} 
        I & J & j' \\ J' & I & 1 
    \end{matrix} \right\} \hat{I}\sqrt{I(I+1)}
    \mathcal{M}_2^{(\rm s.p.)}. \label{MM12}
\end{align}

The single-particle reduced matrix elements are well-known in literature \cite{sha13}. They are decomposed into two components: the geometrical coefficient and the radial overlap integral.
The radial overlap integrals for the $E1$ and $M1$ transitions are different from the $r$-dependence in the operator
\begin{equation}\label{I1}
    \mathcal{I}_{E1} = \int \phi_{n \ell' j'}(r) \chi_{\ell j}(E,r)r \,dr,~~~\mathcal{I}_{M1} = \int \phi_{n \ell' j'}(r) \chi_{\ell j}(E,r) \,dr.
\end{equation}
The potential model for radiative-capture reaction focuses on the calculation of the radial overlap integrals in Eq.~\eqref{I1} with appropriate wave functions, $\chi_{\ell j}$ and $\phi_{n \ell' j'}$.

\subsection{Single-particle wave functions for the overlap integrals}
In our approach, both scattering wave function $ \chi_{\ell j}$ and bound wave function $\phi_{n \ell' j'}$ in Eq.~\eqref{I1} were simultaneously obtained within the Skyrme Hartree-Fock calculation. The detailed Hartree-Fock calculation for bound and scattering single-particle state were presented in Refs. \cite{dov71,dov72}.

The calculation was started with the radial Hartree-Fock equations using the \texttt{skyrme\_rpa} program in Ref.~\cite{col13}
\begin{equation} \label{HFeq}
\left\{\dfrac{\hbar^2}{2m^*(r)}\left[- \frac{d^2}{dr^2} + \dfrac{\ell'(\ell' + 1)}{r^2} \right] + V(r) - \frac{d}{dr}\left[\frac{\hbar^2}{2m^*(r)}\right] \frac{d }{dr} \right\}\varphi_{n \ell' j'}(r) = \varepsilon_{n \ell' j'} \varphi_{n \ell' j'}(r).
\end{equation}
where $m^*(r)$ is the nucleon effective mass in the Skyrme Hartree-Fock formalism. The Skyrme Hartree-Fock potential $V(r)$ consists of the central $V_{\rm c}(r)$, the spin-orbit $V_{\rm s.o.}(r)$, and the one-body Coulomb potential $V_{\rm Coul.}(r)$ in the case of proton
\begin{equation}
    V(r) = V_{\rm c}(r) + qV_{\rm Coul.}(r) + V_{\rm s.o.}(r),
\end{equation}
with $q$ being the charge of the incident nucleon. The SLy4 interaction \cite{cha98} is chosen in our calculation. The first-order derivative term in Eq.~\eqref{HFeq} is eliminated by using the transformation \cite{dov71}
\begin{equation}\label{Dovertrans}
    \varphi_{n \ell' j'}(r) = [m^*_{q}(r)/m]^{1/2}\tilde{\varphi}_\alpha(r),
\end{equation}
with $m$ being the nucleon mass. Then Eq.~\eqref{HFeq} is rewritten as the usual Schr\"odinger equation
\begin{equation}
    \left\{\dfrac{\hbar^2}{2 m'}\left[-\dfrac{d^2}{dr^2} + \dfrac{\ell'(\ell' + 1)}{r^2} \right] + \mathcal{V}_b(\varepsilon_{n \ell' j'}, r) -\varepsilon_{n \ell' j'} \right\} \tilde{\varphi}_{n \ell' j'}(r) = 0, \label{SEb}
\end{equation}
where $m' = m A/(A-1)$ is used for the center-of-mass correction \cite{vau70} that is important for the calculations for light nuclei. The potential $\mathcal{V}_b(\varepsilon_{n \ell' j'}, r)$ in Eq.~\eqref{SEb} is given by
\begin{equation}
    \mathcal{V}_b(\varepsilon_{n \ell' j'}, r) = \mathcal{N}_b \mathcal{V}_b^{\rm c} (\varepsilon_{n \ell' j'},r) + q\mathcal{V}_{\rm Coul.}(r) + \mathcal{V}_{\rm s.o.}(r){\bm{\ell}'}\cdot\bm{\sigma}, \label{mathcalVb}
\end{equation}
where the central part $\mathcal{V}_b^{\rm c} (\varepsilon_{n \ell' j'},r)$ depends on the single-particle energy $\varepsilon_{n \ell' j'}$
\begin{align} \label{pot}
\mathcal{V}_b^{\rm c} (\varepsilon_{n \ell' j'},r) = \dfrac{m^*(r)}{m} \left\{V_{\rm c}(r) + \dfrac{1}{2} \dfrac{d^2}{dr^2}\left(\dfrac{\hbar^2}{2m^*(r)} \right) - \dfrac{m^*(r)}{2\hbar^2} \left[\dfrac{d}{dr}\left(\dfrac{\hbar^2}{2m^*(r)}\right)\right]^2\right\}\nonumber \\
+ \left[1 - \dfrac{m^*(r)}{m}\right] \varepsilon_{n \ell' j'},
\end{align}
and the spin-orbit and Coulomb parts are energy-independent
\begin{align}
   \mathcal{V}_{\rm Coul.}(r) &= [m^*(r)/m] V_{\rm Coul.}(r), \label{Vcoul}\\ 
   \mathcal{V}_{\rm s.o.}(r) &= [m^*(r)/m] V_{\rm s.o.}(r). \label{Vso}
\end{align}

The scattering states of the Skyrme Hartree-Fock field is obtained by replacing $\varepsilon_{n \ell' j'}$ in Eq.~\eqref{SEb} by the continuous energy $E$ \cite{dov71}. The scattering wave function is the solution of
\begin{equation}
\left\{\dfrac{\hbar^2}{2 m'} \left[-\dfrac{d^2}{dr^2} + \dfrac{\ell(\ell + 1)}{r^2} \right] +\mathcal{V}_s(E, r) -E \right\} \chi_{\ell j}(E, r) = 0. \label{SEs}
\end{equation}
The potential $\mathcal{V}_s(E, r)$ in Eq.~\eqref{SEb} is
\begin{equation}
    \mathcal{V}_s(E, r) = \mathcal{N}_s \mathcal{V}_s^{\rm c} (E,r) + q\mathcal{V}_{\rm Coul.}(r) + \mathcal{V}_{\rm s.o.}(r){\bm{\ell}}\cdot\bm{\sigma}. \label{mathcalVs}
\end{equation}
The central part $\mathcal{V}_s^{\rm c} (E,r)$ depends on the scattering energy $E$. It is obtained by replacing the single-particle energy $\varepsilon_{n \ell' j'}$ in Eq.~\eqref{pot} by the positive energy $E$.
The spin-orbit and Coulomb parts are the same as Eqs.~\eqref{Vcoul} and \eqref{Vso} as they are energy independent. Note that two parameters $\mathcal{N}_b$ and $\mathcal{N}_s$ are introduced in Eqs.~\eqref{mathcalVb} and \eqref{mathcalVs}, respectively. 

It is known that the Skyrme Hartree-Fock approximation gives tightly bound single-particle states. The asymptotic behavior of the bound state should be corrected. Therefore, the value of $\mathcal{N}_b$ is adjusted to obtain the single-particle energy $\varepsilon_{n\ell'j'}$ of the first unoccupied state equal to the experimental nucleon separation energy. All occupied single-particle states are kept unchanged. It is well-known as the well-depth method. 

For the scattering state, the mean-field approach such as the Skyrme Hartree-Fock calculation cannot reproduce exactly the resonance state. Therefore, the scattering state has to be corrected to reproduce the exact position of low-lying resonances. It is done by the adjusted value of $\mathcal{N}_s$. 

The single-particle bound and scattering states in Eq.~\eqref{I1} with the corrections have the correct asymptotic forms required by the properties of the low energy radiative-capture reaction. The values of parameters are well-constrained by the experimental values including the nucleon separation energy for $\mathcal{N}_b$ and the energy of the nuclear excited state (or the position of the resonance) for $\mathcal{N}_s$.

At the end of the calculation, the value of $S_F$ is fine-tuned to reproduce the experimental data. The adjustments are typical in the framework of the potential model. The difference in our approach is that the scattering and bound wave functions are initially obtained simultaneously from a microscopic approach such as the Skyrme Hartree-Fock formalism.

Mean-field models are often considered to be inappropriate to describe light nuclei such as $A = 7, 8$ nuclei. There have been works adopting the mean-field procedure \cite{sug96,cha03}. Even for $^4$He some useful conclusions have been drawn using a Skyrme Hartree-Fock calculation \cite{she96} and useful conclusions have be obtained for other light nuclear system \cite{sug96}. In this manuscript we show that a Skyrme Hartree-Fock calculation leads to very reasonable results of radiative-capture cross sections with $A = 7, 8$ nuclei.

\subsection{Partial-wave analysis}
\begin{figure}
    \centering
    \includegraphics[width=0.9\textwidth]{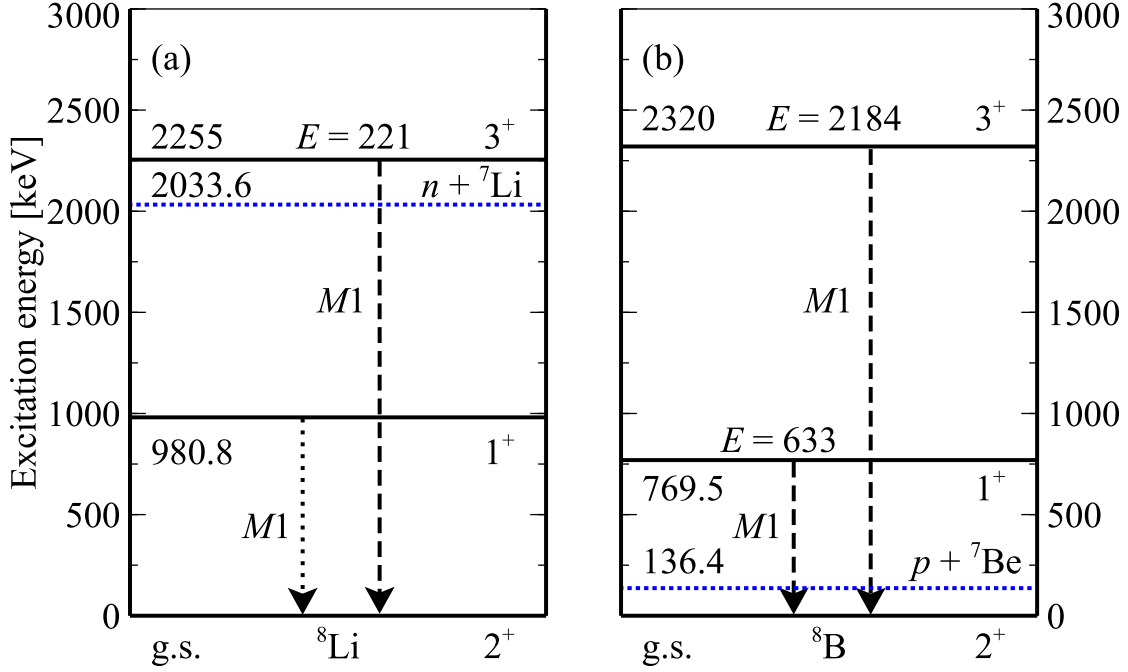}
    \caption{Level schemes of $^8$Li (a) and $^8$B (b). All low-lying resonances are caused by the $M1$ transitions to the ground states ($2^+$) of $^8$Li and $^8$B. The blue dotted lines are the reaction thresholds. The arrows are bound-to-bound $M1$ transitions. The black dashed arrows are the transitions above the threshold. The dotted arrow is the sub-threshold $M1$ transition.}
    \label{lv_scheme}
\end{figure}

Fig.~\ref{lv_scheme} illustrates the level schemes of $^8$Li and $^8$B nuclei with the relevant bound-to-bound $M1$ transitions from the $1^+$ and $3^+$ states to the $2^+$ ground states. Note that two bound-to-bound $M1$ transitions from $3^+$ state to $2^+$ state are similar. The resonance at $2184$ keV of $^7$Be($p,\gamma$)$^8$B reaction and the resonance at $222$ keV of $^{7}$Li($n,\gamma$)$^{8}$Li are related. However, as the radioactive-capture process is the continuum-to-bound transition, they cannot be as similar to each other as the two bound-to-bound $M1$ transitions are. The sub-threshold $M1$ transition (the dotted arrow in Fig.~\ref{lv_scheme}) is not in our calculation.

In the case of the $E1$ transitions, as shown in Table \ref{table 1}, the nucleon is assumed to be captured into the single-particle state $1p_{3/2}$ for both target nuclei. The selected single-particle scattering states are consequently $s$ wave, and $d$ waves including $d_{3/2}$ and $d_{5/2}$. The total spin of the initial state or the channel spin $J^\pi$ are $1^-$ and $2^-$ for the $s$ wave, and $1^-, 2^-$ and $3^-$ for the $d$ waves. Note that the total spin $J$ is unique for the bound-to-bound transition. While the average sum of all possible channel spins is required for the continuum-to-bound transition.
\begin{table}
\setlength{\tabcolsep}{1.0em}
    \centering
    \caption{All $E1$ transitions to the $2^+$ ground states in $^8$Li and $^8$B are considered. $J^\pi$ is the channel spin and parity of the initial system. $\mathcal{N}_s$ and $\mathcal{N}_b$ are the two parameters in the calculation. $S_F$ is the spectroscopic factor. The values of $\mathcal{N}_b$ are determined by the $Q$ values that are $2033$ keV and $136$ keV for $^{7}$Li($n,\gamma$)$^{8}$Li and $^7$Be($p,\gamma$)$^8$B reactions, respectively. 
    }
    \label{table 1}
    \begin{tabular}{ccccccccc}
    \hline \hline
         Reaction & $J^{\pi}$ & $\ell$& $\mathcal{N}_s$ & $n\ell'_{j'}$ & $\mathcal{N}_b$ & $S_F$ \\ \hline
         $^7$Li($n,\gamma$)$^8$Li & $1^{-}$, $2^{-}$ & $s$ & $1.00$ & $1p_{3/2}$ & $0.72$ & $1.00$ \\
          & $1^{-}$, $2^{-}$, $3^{-}$ & $d$ & $1.00$ & $1p_{3/2}$ & $0.72$ & $1.00$ \\
         $^7$Be($p,\gamma$)$^8$B & $1^{-}$, $2^{-}$ & $s$ & $1.00$ & $1p_{3/2}$ & $0.70$ & $0.80$ \\
          & $1^{-}$, $2^{-}$, $3^{-}$ & $d$ & $1.00$ & $1p_{3/2}$ & $0.70$ & $0.80$ \\
    \hline \hline
    \end{tabular}
\end{table}

For the $M1$ transitions, there are the resonances corresponding to the bound-to-bound transitions illustrated by the dashed arrows in Fig.~\ref{lv_scheme}. 
The resonance at $633$ keV is caused by the proton in the $p$-scattering wave captured into the $1p_{3/2}$-bound state. The channel spins $J^\pi$ are $1^+$, $2^+$ and $3^+$ for the $p$ wave with $j = 3/2$, but there is no $J^\pi = 3^+$ for $j = 1/2$. The configuration is the same for the case of $M1$ resonance at 222 keV in $^7$Li($n, \gamma$)$^8$Li reaction. For the resonance at $2184$ keV  of $^7$Be($p, \gamma$)$^8$B reaction, the $d$ wave is dominant that is discussed in the next section. Table \ref{table 2} presents the configurations in the calculation for the $M1$ transitions.
\begin{table}
\setlength{\tabcolsep}{0.9em}
    \centering
    \caption{ 
    The same as Table \ref{table 1} but for the $M1$ transitions. The energies (in keV) of the $M1$ resonances are shown. Note that there is no $J^\pi = 3^+$ for $p_{1/2}$ states.}
    \label{table 2}
    \begin{tabular}{cccccccccc}
    \hline \hline
         Reaction & $E$ [keV] & $J^{\pi}$ & $\ell$& $\mathcal{N}_s$ & $n\ell'_{j'}$ & $\mathcal{N}_b$ & $S_F$ \\ \hline
         $^{7}$Be($p,\gamma$)$^{8}$B & $633$ & $1^{+}, 2^{+}, 3^{+}$ & $p$ & $0.67$ & $1p_{3/2}$ & $0.70$ & $1.00$ \\
          & $2184$ & $1^{+},2^{+},3^{+}$ & $d$ & $1.27$ & $1d_{5/2}$ & $1.44$ & $0.10$ \\
         $^{7}$Li($n,\gamma$)$^{8}$Li & $222$ & $1^{+},2^{+},3^{+}$ & $p$ & $0.83$ & $1p_{3/2}$ & $0.58$ & $0.05$ \\
    \hline \hline
    \end{tabular}
\end{table}

\section{RESULTS AND DISCUSSIONS}
\subsection{Non-resonant $E1$ transition in $^{7}$Li($n,\gamma$)$^{8}$Li and $^{7}$Be($p,\gamma$)$^{8}$B reaction}

As the $^{7}$Li target is not radioactive, the $^{7}$Li($n,\gamma$)$^{8}$Li reaction was sufficiently studied for the constraint of the optical parameters used in the case of the $^{7}$Be($p,\gamma$)$^{8}$B reaction \cite{des94,bro96,ben99,nag05}. Both reactions are analyzed in the study to support the consistency of the approach. The parameters used for the $E1$ transition are shown in Table \ref{table 1}.

\begin{figure}
    \centering
    \includegraphics[width=0.9\textwidth]{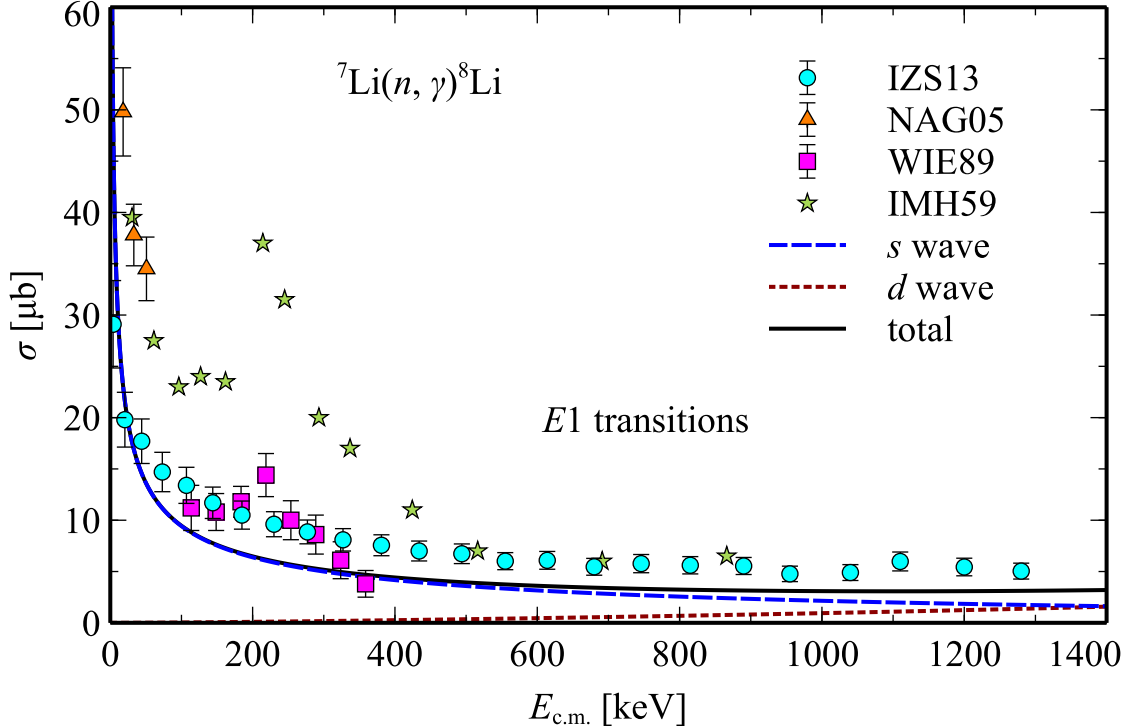}
    \caption{The total $E1$ transition (the solid line) caused by $s$ (the dashed line) and $d$ waves (the dotted line) in the partial-wave analysis. $S_F = 1.0$. The experimental data were taken from Refs.~\cite{izs13,nag05,wie89,imh59}.}
    \label{7Li_E1}
\end{figure}
For the $^{7}$Li($n,\gamma$)$^{8}$Li reaction, the cross sections were measured over a wide range of energies from a few keV up to 1000 keV \cite{imh59,wie89,nag05}. In our calculation, the results in Fig.~\ref{7Li_E1} show that the main contribution comes from the $s$ wave. The experimental data are well reproduced with only the adjustment for the single-particle bound state. The parameter $\mathcal{N}_b$ is equal to $0.72$ to reproduce the neutron separation energy $2033$ keV in $^8$Li. As there is no $E1$ resonance in the energy region from the threshold up to $1400$ keV for the calibration, the parameters $\mathcal{N}_s$ for all scattering states are kept unity. 

The $S_F$ in this case is not required for the adjustment. The work in Ref.~\cite{izs13} has reported on the Coulomb dissociation of $^8$Li in order to determine the neutron-capture cross section for the inverse reaction $^7$Li($n,\gamma$)$^8$Li. Our calculation gives the calculated $E1$ curve passes through the experimental data points in Ref.~\cite{izs13} that are lower than those in Ref.~\cite{nag05} under 200 keV. Around the resonance of 222 keV, our calculation for the $E1$ transition passes through the data points in Ref.~\cite{wie89} (Fig.~\ref{7Li_E1}). The shell-model calculations in Refs.~\cite{tsa05,tim13} gave $S_F$ to be $0.62$ and $1.143$, respectively. In addition, the $S_F$ in the \textit{ab initio} calculation is $0.966$ \cite{tim13}.

\begin{figure}
    \centering
    \includegraphics[width=0.9\textwidth]{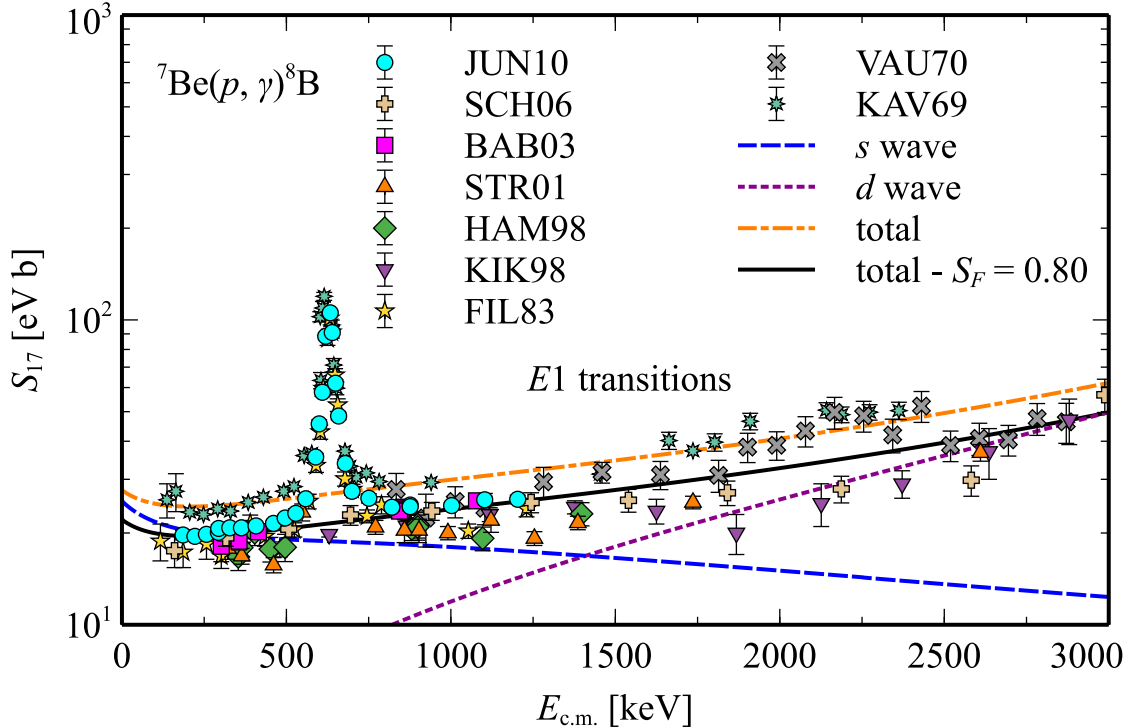}
    \caption{The $E1$ transitions caused by $s$ and $d$ waves in the partial-wave analysis. The experimental data were taken from Refs.~\cite{kav69,vau70,fil83,kik98,ham98,str01,sch06,bab03,jun10}.}
    \label{7Be_E1}
\end{figure}

Table~\ref{table 1} shows the consistencies between $^{7}$Li($n,\gamma$)$^{8}$Li and $^7$Be($p,\gamma$)$^8$B reaction. The identical initial states and analyses make their results similar. The only considerable difference between the two reactions is that the experimental separation energy is $136$ keV for proton \cite{wan12} and $2033$ keV for neutron obviously because of the Coulomb energy. However, the parameter $\mathcal{N}_b$ in the case of the $^7$Be target is equal to $0.70$ that is close to the case of $^7$Li target as shown in Table \ref{table 1}. The partial-wave analysis of $E1$ transitions in $^7$Be($p,\gamma$)$^8$B reaction is presented in Fig.~\ref{7Be_E1}. The parameters $\mathcal{N}_s$ for the scattering states are also kept unity. The $S_F$ is 0.8 for this case to reproduce the experimental data point in Ref.~\cite{jun10}. The values of $0.742$ and $1.079$ were reported in the shell-model calculations \cite{tim13}, while $S_F = 0.884$ in the \textit{ab initio} calculation \cite{tim13}. The three-cluster approach in Ref.~\cite{des04PRC} gave $S_F = 0.836$. 

In terms of nuclear astrophysics, the cross section at zero energy of the $^{7}$Be($p,\gamma$)$^{8}$B reaction known as $\mathcal{S}_{17}(0)$ is of particular interest. As shown in many previous works, it is determined entirely by the $E1$ transition. The contribution of the resonant $M1$ transition can be negligible.
Table~\ref{table 3} shows the values of $\mathcal{S}_{17}(0)$ reported in the wide range of both experiments and theoretical calculations taken from previous works. Considering only the $E1$ transition, $\mathcal{S}_{17}(0) = 19.1_{-1.0}^{+4.0}$ and $19.4$ eV\,b were given in Refs.~\cite{cha03} and \cite{nav11}, respectively. Within the shell-model calculation, the value of $19.424$ eV\,b was reported in Ref.~\cite{ben99}. From the study of $^7$Li($d,p$)$^8$Li cross section, $\mathcal{S}_{17}(0)$ in Refs.~\cite{par66}, \cite{kav69} and \cite{vau70} were $25.8\pm 2.2$, $24.3\pm 2.0$ and $17.4 \pm 1.6$ eV\,b, respectively. The values of $18.5 \pm 1.7$, $18.4 \pm 1.6$, and $18.6 \pm 0.4$ eV\,b were reported in Refs.~\cite{ham98}, \cite{str01} and \cite{dav03}, respectively. Most recently, the experimental values of $20.6\pm 0.8$ and $21.5\pm 0.6$ eV\,b were given in Refs.~\cite{sch06} and \cite{jun10}, respectively. Theoretically, Refs.~\cite{tur21} and \cite{kra22} have recently given the values of $20.51^{+2.02}_{-1.85}$ and $19.8\pm 0.3$ eV\,b, respectively.
\begin{table}[b]
\setlength{\tabcolsep}{0.9em}
    \centering
    \caption{The selected values of the $\mathcal{S}_{17}(0)$ taken from experimental and theoretical studies.}
    \label{table 3}
    \begin{tabular}{lc}
    \hline \hline
         Experiment  & $\mathcal{S}_{17}(0)$ [eV\,b] \\ \hline
         Parker (1966) \cite{par66} & $25.8\pm 2.2$ \\
         Kavanagh \textit{et al.} (1969) \cite{kav69} & $24.3\pm 2.0$ \\
         Vaughn \textit{et al.} (1970) \cite{vau70} & $17.4 \pm 1.6$ \\
         Hammache \textit{et al.} (1998) \cite{ham98} & $18.5 \pm 1.7$ \\
         Strieder \textit{et al.} (2001) \cite{str01} & $18.4\pm 1.6$ \\ 
         Davids and Typel (2003) \cite{dav03} & $18.6\pm 0.4$ \\ 
         Sch\"umann \textit{et al.} (2006) \cite{sch06} & $20.6\pm 0.8$ \\
         Junghans \textit{et al.} (2010) \cite{jun10} & $21.5\pm 0.6$ \\ \hline
         Theory  & $\mathcal{S}_{17}(0)$ [eV\,b] \\ \hline
         Bennaceur \textit{et al.} (1999) \cite{ben99} & $19.424$ \\
         Chandel \textit{et al.} (2003) \cite{cha03} & $19.1_{-1.0}^{+4.0}$ \\
         Navr\'atil \textit{et al.} (2011) \cite{nav11} & $19.4$ \\
         Tursunov \textit{et al.} (2021) \cite{tur21} & $20.51^{+2.02}_{-1.85}$ \\
         Kravvaris \textit{et al.} (2022) \cite{kra22} & $19.8\pm 0.3$ \\
         Present work & $22.3$ \\
    \hline \hline
    \end{tabular}
\end{table}

\begin{figure}[t]
    \centering
    \includegraphics[width=0.9\textwidth]{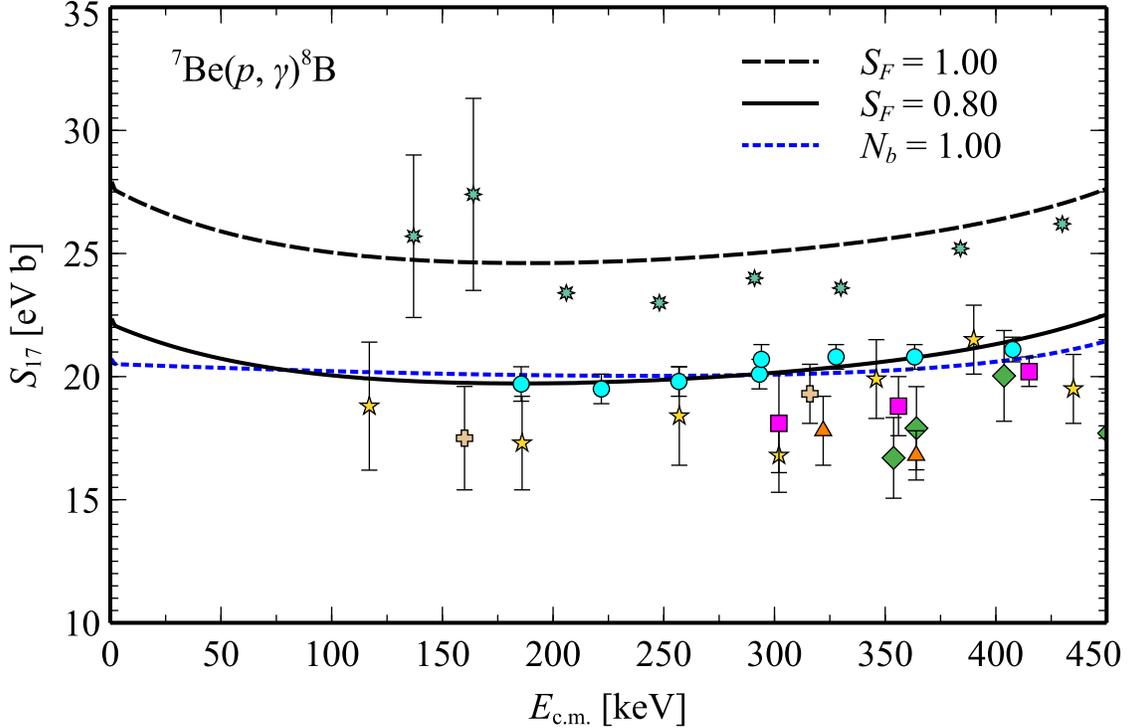}
    \caption{The relevance of $\mathcal{S}_{17}$ factor of $^7$Be($p,\gamma$)$^8$Be at low energy. The calculation using $\mathcal{N}_b = 0.70$ gives $\mathcal{S}_{17}(0) = 22.3$ eV\,b. While the values of $21.2$ eV\,b is given by calculation with $\mathcal{N}_b =1.00$. The experimental data are from JUN10 \cite{jun10} (the circle), SCH06 \cite{sch06} (the cross), BAB03 \cite{bab03} (the square), STR01 \cite{str01} (the triangle), HAM98 \cite{ham98} (the diamond), FIL83 \cite{fil83} (the pentagram), and KAV69 \cite{kav69} (the octagram).}
    \label{7Be_ext}
\end{figure}

Our calculation with the only $E1$ transition being considered and $S_F = 1.0$ gives the $\mathcal{S}_{17}(0) = 27.8$ eV\,b that is the upper limit in our calculation (Fig.~\ref{7Be_ext}). It is $22.2$ eV\,b using the value $S_F = 0.8$ fine-tuned to reproduce the experimental data points in Ref.~\cite{jun10}. The results pointed out that the $E1$ transitions are well-described even in the case without any low-lying resonance for the calibration. The parameters $\mathcal{N}_s$ for the scattering state are unity in all cases of non-resonant $E1$ transitions in the study. The only adjustment for the $E1$ transition is the parameter $\mathcal{N}_b$. The value of $S_{17}(0)$ in our calculation does not strongly depend on the value of the parameter $\mathcal{N}_b$. The calculation with $\mathcal{N}_b = 1.00$ is also shown in Fig.~\ref{7Be_ext} as the dotted line. The value of $S_{17}(0)$ is $21.2$ eV\,b. $S_F = 0.08$ is required to reproduce the experimental data. The inadequate value of $S_F$ is well-understood. The single-particle state obtained in the Hartree-Fock calculation is tightly bound. While the radiative capture reaction at low energy requires a weakly bound system. The value of $S_{17}(0)$ is, however, not much changed.

\subsection{Low-lying resonant $M1$ transition in $^{7}$Li($n,\gamma$)$^{8}$Li and $^{7}$Be($p,\gamma$)$^{8}$B reaction}
The potential model is sufficient not only for the $E1$ transition but also for the $M1$ transition. In the study, the $M1$ transitions include resonances that are at $633$ keV and $2184$ keV in the case of the $^{7}$Be($p,\gamma$)$^{8}$B reaction and at $222$ keV in $^{7}$Li($n,\gamma$)$^{8}$Li reaction. The parameters $\mathcal{N}_s$ for the $M1$ transition are adjusted to reproduce the position of the resonance.

\begin{figure}[t]
    \centering
    \includegraphics[width=0.9\textwidth]{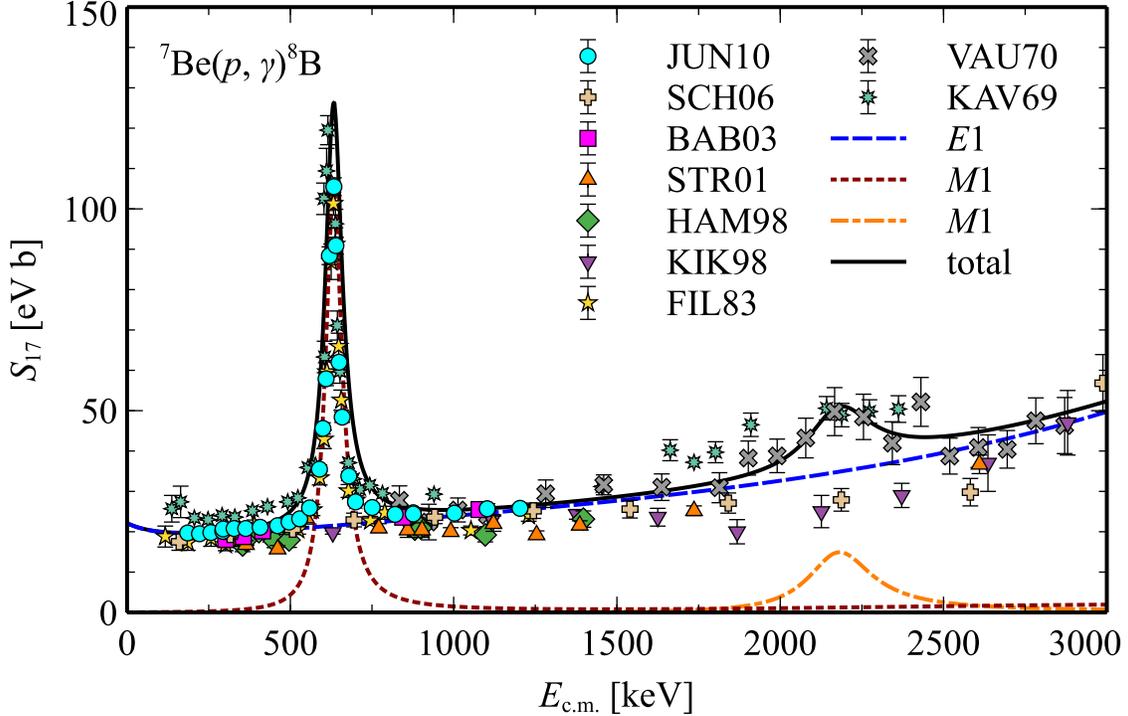}
    \caption{All transitions are considered. The dashed line is the contribution of the $E1$ transition as shown in Fig.~\ref{7Be_E1}. The dotted and dash-dotted lines are of the $M1$ transitions. The solid line shows our final result. The experimental data were taken from Refs.~\cite{kav69,vau70,fil83,kik98,ham98,str01,bab03,sch06,jun10}.}
    \label{7Be_All}
\end{figure}

First, in the case of the $^{7}$Be($p,\gamma$)$^{8}$B reaction, as in our analysis, the $M1$ transitions from the $p$-scattering states to the $1p_{3/2}$-bound state are responsible for the first resonance at $633$ keV. The same discussion was given in Refs.~\cite{rob73,ber96}. The parameters $\mathcal{N}_s$ were adjusted to reproduce the correct position of the resonance (see Table~\ref{table 2}). Previous theoretical works concluded that $S_F$ should be unity for this case \cite{kim87,ber96,tur21}. We also use $S_F = 1.0$ in our calculation. The shape of the resonance is well reproduced. As shown in Fig.~\ref{7Be_All}, the contribution of the $M1$ transition is negligible when the energy is below 400 keV.

The second resonance at $2184$ keV corresponding to the excited state ($3^+$) of $^8$B is caused by $M1$ transition to the ground state ($2^+$) of $^8$B \cite{sel84,til04}. The work in Ref.~\cite{tur21} proposed that the $E2$ transition from scattering $d$ waves is responsible for this resonance. However, the known level scheme illustrated in Fig.~\ref{lv_scheme} shows only the bound-to-bound $M1$ transition. In our calculation, the $d$-scattering wave is also adopted but using the $M1$ instead of the $E2$ transition. Following this argument, there should be another bound configuration to reproduce the resonant $M1$ transition. Our assumption is that a proton is captured into the $1d_{5/2}$ state from $d$-scattering waves with the channel spins $J^\pi = 1^+, 2^+$, and $3^+$. The $\mathcal{N}_b = 1.44$ to reproduce the proton separation energy of $136$ keV for this case. The parameters $\mathcal{N}_s$ are listed in Table~\ref{table 2} to calibrate the position of resonance caused by $d$-scattering waves. The result of the calculations for the second resonance is shown in Fig.~\ref{7Be_All}. The $S_F$, in this case, is $0.1$. Its small value can be interpreted that the interference of other configurations appearing at this energy. As much high incident energy as much small $S_F$ is obtained. Besides, at the energy $2988$ keV, the study in Ref.~\cite{tur21} analyzed the resonant $E1$ transitions from $d$-scattering waves to reproduce experimental data taken from Ref.~\cite{sch06}. This resonance possibly affects the analysis of the second resonance at $2184$ keV. However, the experimental data supporting this resonance is not sufficient. Therefore, it is not taken into account in our analysis in the present work.
The total $\mathcal{S}_{17}$ factor of $^{7}$Be($p,\gamma$)$^{8}$B reaction with the $E1$ and $M1$ transitions is displayed in Fig.~\ref{7Be_All} with our full calculation. As shown in Fig.~\ref{7Be_ext} and Table~\ref{table 3}, the $\mathcal{S}_{17}(0)$ in our calculation including $M1$ transitions is now $22.3$ eV\,b. 

\begin{figure}[t]
    \centering
    \includegraphics[width=0.9\textwidth]{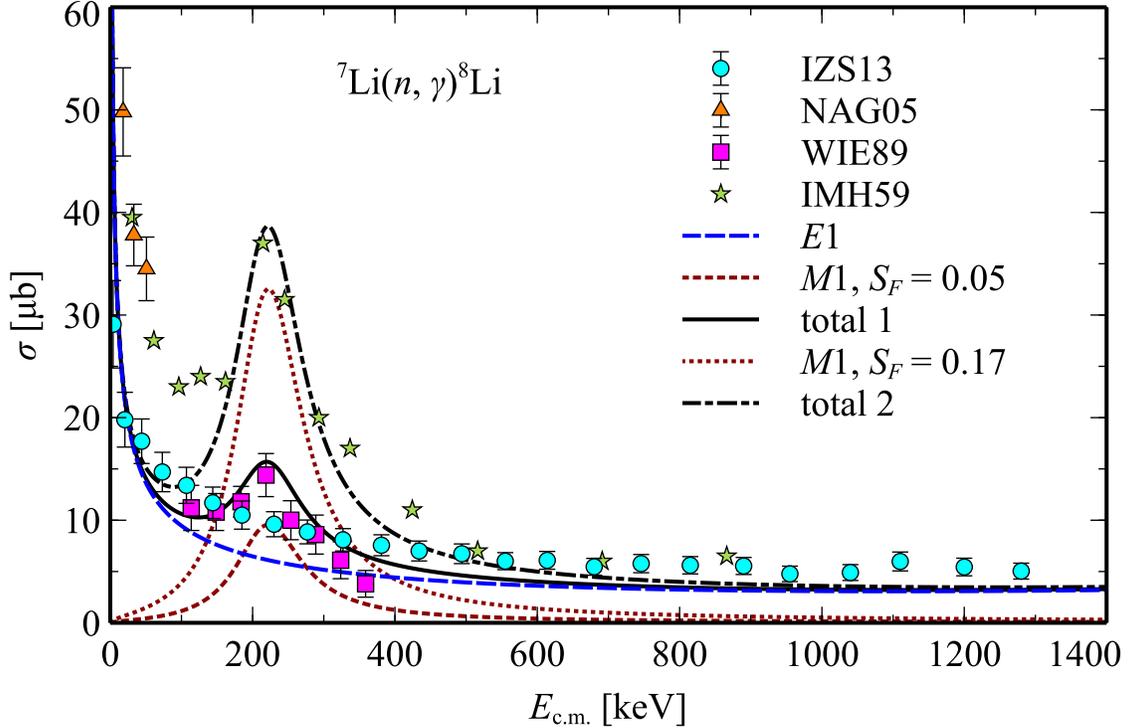}
    \caption{The calculation for $^{7}$Li($n,\gamma$)$^{8}$Li radiative-capture reaction. All electromagnetic transitions are in consideration. The values of $S_F$ is adjusted to reproduce the data of Ref.~\cite{imh59} ($S_F = 0.17$) or Ref.~\cite{wie89} ($S_F = 0.05$).}
    \label{7Li_All}
\end{figure}

Analogously, the calculation is applied to the resonant $M1$ transition at $222$ keV in $^{7}$Li($n,\gamma$)$^{8}$Li reaction (Fig.~\ref{7Li_All}). The scattering $p$-scattering waves captured into $1p_{3/2}$ bound state via the $M1$ transition are the source of the resonance. The cross sections measured at $222$ keV in Refs.~\cite{imh59} and \cite{wie89} are different. The value of $S_F$ in our calculation is $0.17$ or $0.05$ in order to reproduce the data given by Ref.~\cite{imh59} or \cite{wie89}, respectively. As the resonances at $2184$ keV and at $222$ keV are related, one can expect that the value of $S_F$ in this case is at the same order of the one obtained from the calculation for the resonance at $2184$ keV above. However, as the $3^+$ state in $^8$Li is closed to the threshold of $n+^7$Li, $S_F$ in this case should be significantly reduced. That is the reason why $S_F = 0.05$ is chosen to reproduce the experimental data in Ref.~\cite{wie89}.

\begin{figure}[b]
    \centering
    \includegraphics[width=0.9\textwidth]{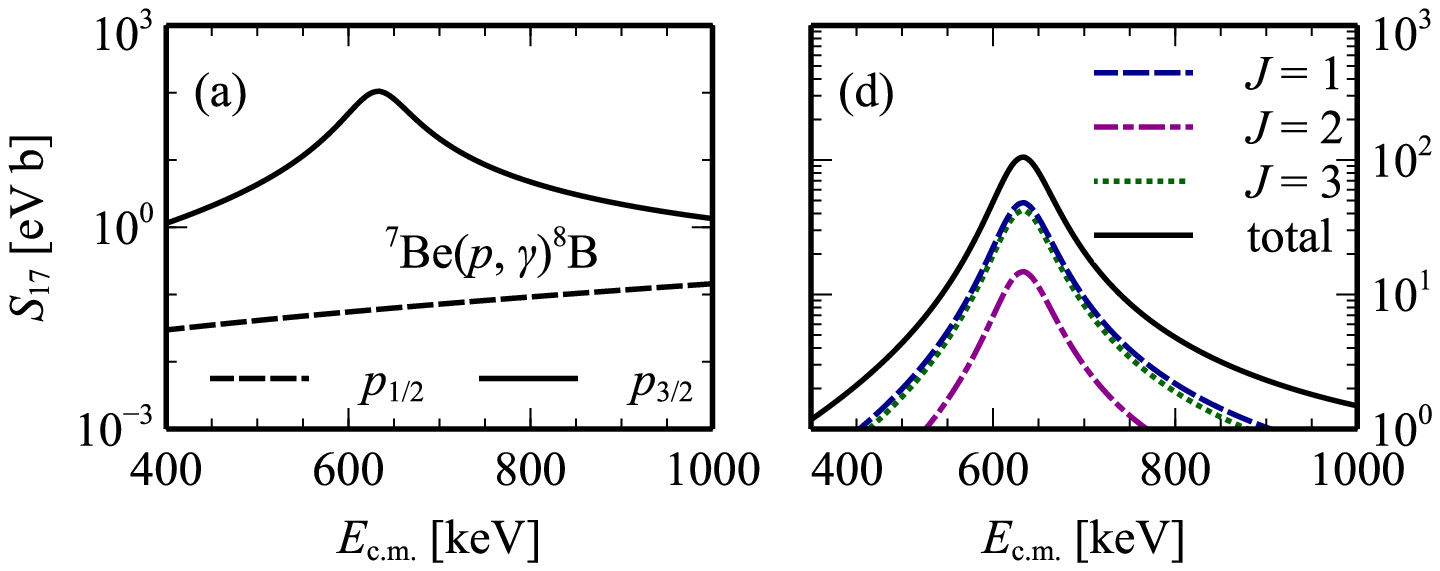}
    \includegraphics[width=0.9\textwidth]{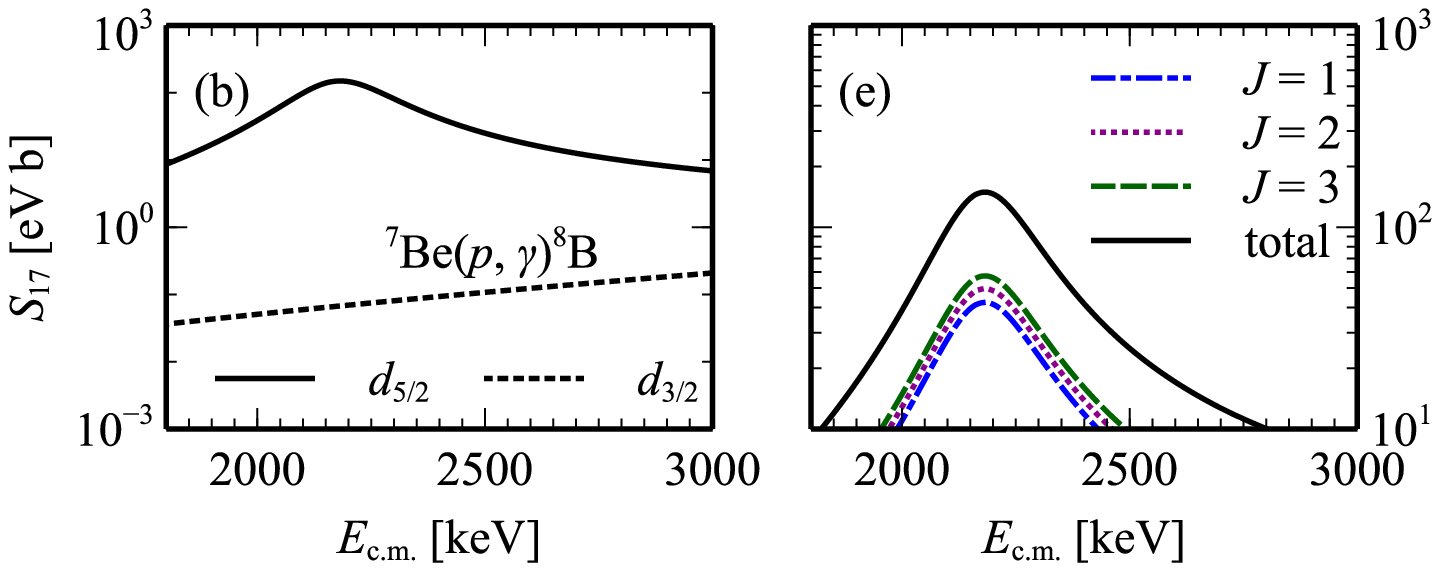}
    \includegraphics[width=0.9\textwidth]{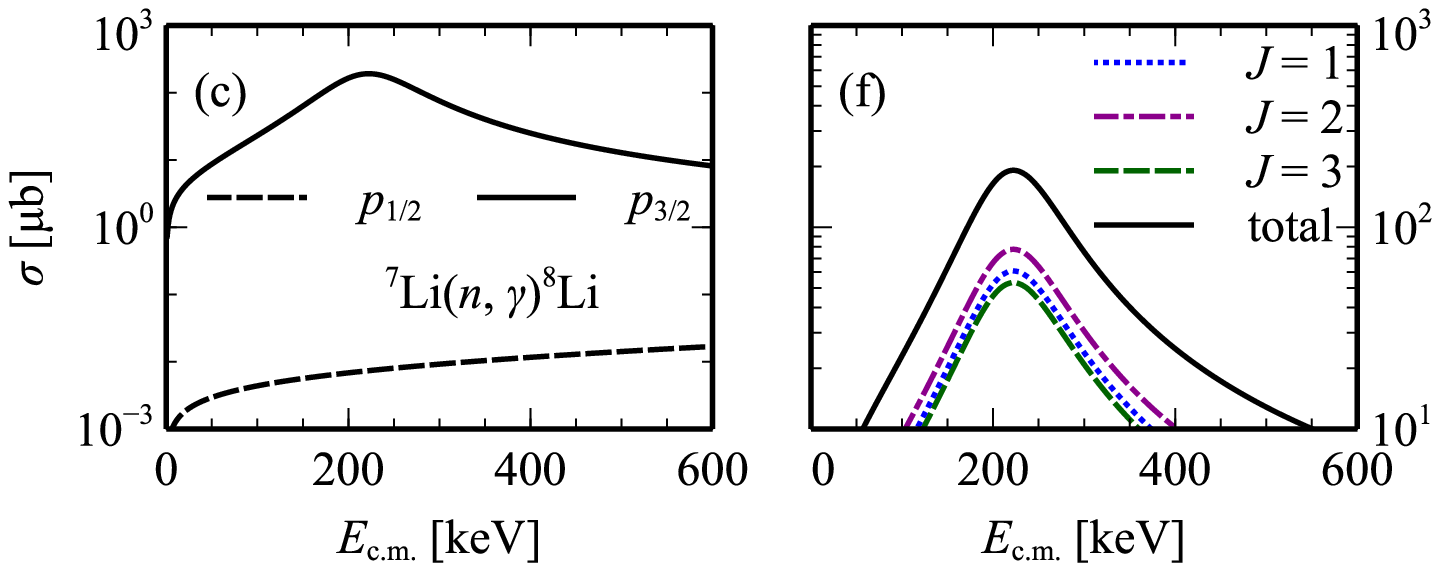}
    \caption{The partial analysis for the three $M1$ resonances. In Figs. (a), (b), and (c), the contributions by the spin-orbit partners with all channel spin are shown. Figs. (d), (e), and (f) illustrate the contribution of different channel spins.}
    \label{partialanalysis}
\end{figure}

Note that only one spin-orbit partner of the partial waves is the source of the resonance due to the spin-orbit potential included in our calculation. The spin-orbit partner waves, $1p_{1/2}$ and $1p_{3/2}$ for an example, shares the same parameter $\mathcal{N}_s$ (Table \ref{table 2}). Without the spin-orbit potential, both $p$ waves can contribute to the resonance. The spin-orbit potential should not be neglected as its especially important effect in light nuclei was emphasized in Ref.~\cite{dov71,cha03}. The contributions of different scattering waves are shown in Figs.~\ref{partialanalysis}(a-c) for the three resonances.

It is worthwhile to remind that the channel spin $J$ made the largest contribution can not be determined by the $J$ of the corresponding nuclear-structure state. The contributions of different channel spins in our calculation are shown in Figs.~\ref{partialanalysis}(d-f). As shown in Table \ref{table 2}, the scattering waves causing the resonances are different. While the resonance at $2184$ keV is caused by the $d$-scattering wave, the $p$-scattering wave is the source of the resonance at $222$ keV. This difference leads to the following consequence. In the case of the resonance at $2184$ keV, Fig.~\ref{partialanalysis}(d) shows that the largest contribution is from $J = 3$. Meanwhile, in the case of the resonance at $222$ keV, $J = 3$ shown in Fig.~\ref{partialanalysis}(f) has the smallest contribution due to the absence of the $p_{1/2}$ wave.

\section{Conclusions}
The Skyrme Hartree-Fock potential model is a good approach for the radiative-capture reactions at very low energy for light nuclei. The low-energy scattering problem is comfortably overcome by the Hartree-Fock calculation in the continuum. In addition the bound state is established on the traditional Hartree-Fock formalism and the well-depth method. The calculations for the $E1$ transitions well reproduced the experimental data with only one parameter. The approach is also sufficient to reproduce consistently all $M1$ transitions in the study. The value of the $S_F$ that is complicated to be obtained exactly within the potential model can be determined by the precise measurements at a few hundred-keV energy that is accessible by the experiments \cite{jun10, wie89}. In the case of the $^7$Be($p,\gamma$)$^8$B reaction, the $\mathcal{S}_{17}(0)$ factor determined by the $E1$ transition is $22.3$ eV\,b. The value is not strongly dependent on the parameter in the calculation.

\section*{Acknowledgments}
N.L.A. acknowledges the support by Ho Chi Minh City University of Education Foundation for Science and Technology under grant number CS.2021.19.45. This work was supported by the Institute for Basic Science (IBS-R031-D1).

\bibliography{refs}
\end{document}